\documentclass[aps,prl,preprint, titlepage,showpacs]{revtex4}

\usepackage{graphicx}

\usepackage{dcolumn}

\usepackage{bm}

\begin{document}

\title{Shear viscosity of two-dimensional Yukawa systems in liquid state}

\author{Bin Liu}
\email{bliu@newton.physics.uiowa.edu}

\author{J. Goree}
\email{john-goree@uiowa.edu}

\affiliation{Department of Physics and Astronomy, The University
of Iowa, Iowa City, Iowa 52242}

\date{\today}

\begin{abstract}
The shear viscosity of a two-dimensional (2D) liquid was
calculated using equilibrium molecular dynamics simulations with a
Yukawa potential. The shear viscosity has a minimum, at a Coulomb
coupling parameter $\Gamma$ of about 17, arising from the
temperature dependence of the kinetic and potential contributions.
Previous calculations of 2D viscosity were less extensive, and for
a different potential. The stress autocorrelation function was
found to decay rapidly, contrary to some earlier work. These
results are useful for 2D condensed matter systems and are
compared to a recent dusty plasma experiment.

\end{abstract}

\narrowtext

\maketitle

Two-dimensional systems in crystalline or liquid
states~\cite{Strandburg88} are of interest in various fields of
physics. Monolayer particle suspensions can be formed in colloidal
suspensions~\cite{Keim04} and dusty plasmas~\cite{Melzer00}.
Electrons on the surface of liquid helium form a 2D Wigner
crystal~\cite{Grimes79}. Ions in a Penning trap can be confined
as a single layer of a one-component plasma
(OCP)~\cite{Mitchell99}. Magnetic flux lines in 2D
high-temperature superconductors form patterns of hexagonally
correlated vortices~\cite{Gammel87}. At an atomic scale, gas
atoms adsorb on the surface of substrates such as
graphite~\cite{Stephens79}. Here we will be concerned with
liquids, including liquids near freezing, composed of molecules or
particles that interact with a Yukawa pair potential.

The Yukawa pair potential is widely used in several fields. These
include colloids, monolayer strongly-coupled dusty plasmas, and
some polyelectrolytes~\cite{Denton03} in biological and chemical
systems. The Yukawa potential energy for two particles of charge
$Q$ separated by a distance $r$ is
$U(r)=Q^{2}(4\pi\epsilon_{0}r)^{-1}\exp(-r/\lambda_{D})$, where
$\lambda_{D}$ is a screening length. This potential changes
gradually from a long-range $r^{-1}$ Coulomb repulsion to a
hard-sphere-like repulsion as the screening parameter
$\kappa=a/\lambda_{D}$ is increased. Here, $a=(n\pi)^{-1/2}$ is
the 2D Wigner-Seitz radius~\cite{Kalman04} and $n$ is the areal
number density of particles.

The literature has only a few reports of molecular dynamics (MD)
simulations for computing the shear viscosity of 2D
liquids~\cite{Hoover95,Gravina95}, and none of those are for a
Yukawa potential. Here we present such a simulation, yielding
results for the shear viscosity and the shear stress
autocorrelation function (SACF). We will compare results to a
recent 2D experiment~\cite{Vova04} and to MD
simulations~\cite{Donko00, Sanbonmatsu01,Saigo02, Salin02} for
the shear viscosity of 3D liquids.

Our first motivation arises from the need to model the recent
dusty plasma experiment of Nosenko and Goree~\cite{Vova04}. The
experiment was performed with a monolayer of polymer microspheres
suspended in a plasma. The microspheres interacted with a Yukawa
pair potential~\cite{Konopka00}. In an undisturbed state,
particles arranged themselves in a 2D triangular lattice, which
was then melted by an externally-applied velocity shear due to two
counterpropagating laser beams applied in situ. In this way, a 2D
liquid was produced that had a shear flow. The experimenters
measured $\eta$ and found its variation with temperature.

Our second motivation is to compare $\eta$ for 2D and 3D liquids,
both with a Yukawa potential. Because shear viscosity has the
different units of kg m$^{-1}$s$^{-1}$ and kg s$^{-1}$ in 3D and
2D, respectively, we divide by the volume and areal mass density
respectively, yielding the kinematic viscosity $\nu$. This
quantity has the same units of m$^{2}$s$^{-1}$ for both 3D and 2D,
thereby allowing a comparison of results for 3D and 2D.

Our simulation uses an equilibrium method to calculate $\eta$.
Under equilibrium conditions, momentum transport arises from
random thermal fluctuations in a homogeneous sample, and there is
no macroscopic shear flow. In contrast, we note that in a shear
flow the system is not in equilibrium, and the viscosity often
depends on the applied shear rate.

In an equilibrium method, shear viscosity can be calculated using
the Green-Kubo relation~\cite{Hansen86}. Green-Kubo formulae in
general yield a macroscopic phenomenological transport
coefficient, such as the viscosity or diffusion coefficient, which
is written as a time integral of a microscopic time-correlation
function. The Green-Kubo approach is based on a Liouville
description of a fluid assuming that microscopic fluctuations are
linear and the system has no nonequilibrium fields. Green-Kubo
formulae also assume the validity of the Onsager hypothesis, i.e.,
that spontaneous fluctuations in microscopic quantities decay
according to hydrodynamic laws, and that hydrodynamic quantities
are meaningful. This requires that time scales are long compared
to the collision time and that the system size is large compared
to the mean free path.

To compute the shear viscosity, we start with time series data for
the positions $(x_{i},y_{i})$ and velocities $(v_{x,i}, v_{y,i})$
of $N$ particles, as well as the shear stress
\begin{equation}\label{stress}
 P^{xy}(t)=\sum_{i=1}^{N}mv_{x,i}v_{y,i}-\sum_{i}\sum_{j>i}\frac{x_{ij}y_{ij}U'(r_{ij})}{r_{ij}}.
\end{equation}
The first term of Eq.~(\ref{stress}) is a kinetic part, which
depends only on particle velocities, and the second term is a
potential part, which depends on the pair potential. Here $m$ is
the particle mass and $\mathbf{r}_{ij}=(x_{ij},y_{ij})$ is the
distance between particles $i$ and $j$. We can then compute the
shear stress autocorrelation function (SACF)
\begin{equation}\label{stresscorr}
C_{shear}(t)=\langle P^{xy}(t)P^{xy}(0)\rangle.
\end{equation}
Finally, we find $\eta$ by integrating the SACF using the
Green-Kubo relation
\begin{equation}\label{Kubo}
  \eta=\frac{1}{Ak_{B}T}\int_{0}^{\infty}C_{shear}(t)dt,
\end{equation}
for a 2D liquid of area $A$ and temperature $T$. Equation
(\ref{Kubo}) yields the hydrodynamic parameter $\eta$ based on
fluctuating microscopic parameters entering into the shear stress
$P^{xy}(t)$.

We use normalized units in this letter. The length and time are
normalized by $a$ and $\omega_{pd}^{-1}$, respectively, where
$\omega_{pd}=(Q^{2}/2\pi\epsilon_{0}ma^{3})^{1/2}$~\cite{Kalman04}.
The normalized temperature is $\Gamma^{-1}$, where
$\Gamma=Q^{2}/4\pi\epsilon_{0}akT$ is the Coulomb coupling
parameter, so that a high temperature corresponds to a small
$\Gamma$. The 2D viscosity $\eta$ is normalized by
$\eta_{0}=nm\omega_{pd}a^{2}$, and the kinematic viscosity
$\nu=\eta/nm$ is normalized by $\omega_{pd}a^{2}$.

We performed an MD simulation to calculate $\eta$. The equations
of motion for $N$ particles were integrated using periodic
boundary conditions. A thermostat was applied to achieve a
constant $T$. We recorded particle positions and velocities, and
we used Eqs.~(\ref{stress})-(\ref{Kubo}) to calculate $\eta$.

Our simulation model resembles the experimental system in
Ref.~\cite{Vova04}. In both of them, particles in a monolayer
interact with a Yukawa potential. The values of $\kappa$ and
$\Gamma$ were similar; all our simulations were performed for
$\kappa=0.56$ while the experiment had $\kappa=0.53$. There are,
however, significant differences. The simulation is for
equilibrium conditions, while the experiment of
Ref.~\cite{Vova04}, like most experiments to measure $\eta$, used
an externally-applied shear and therefore resulted in a
measurement of $\eta$ under nonequilibrium conditions. The
simulation had periodic boundary conditions, unlike the
experiment, and the equation of motion when a thermostat is used
does not explicitly model the frictional damping of particle
motion due to gas in the experiment.

We now review the details and tests of our simulation method. We
used a velocity Verlet integrator~\cite{Toxvaerd91} with a time
step $0.02<\Delta t<0.05$ $\omega_{pd}^{-1}$. We verified that
$\Delta t$ was small enough by performing a test, with no
thermostat, where we required a fluctuation of total energy $<3\%$
over an interval of $750~\omega_{pd}^{-1}$. We truncated the
Yukawa potential at $r_{cut}=22a$, with a switching function to
give a smooth cutoff between $20a\leq r\leq22a$. We verified that
the potential energy of the entire system was almost independent
of $r_{cut}$, for $r_{cut}>12a$. We used $N=1024$ particles,
corresponding to a rectangle $56.99a\times49.08a$. The size of
this simulation box limits the maximum meaningful time for
correlation functions to 46~$\omega_{pd}^{-1}$, computed as the
time for a compressional sound wave to transit the box. Later we
will find that except for $\Gamma>124$, which is near freezing,
the SACF decays to zero in less than 46 $\omega_{pd}^{-1}$,
indicating that our simulation box was chosen sufficient small.
Ewald summation was not used because the simulation box was wider
than $\lambda_{D}$ by a factor of 27. The ratio of the two sides
of the box were chosen to allow a perfect triangular lattice to
form at high $\Gamma$, i.e., at low temperature.

After completing these tests, we added the Nos\'{e}-Hoover
thermostat~\cite{Hoover85} to the equation of motion. We tested
the thermostat with different values of the thermal relaxation
time, and we chose a value of 1.0~$\omega_{pd}^{-1}$, which
resulted in a canonical distribution within a time
$4000~\omega_{pd}^{-1}$. To verify that a thermal equilibrium was
attained, we performed the customary test~\cite{Holian95} of
temperature fluctuations, characterized by their variance and
skewness. We also verified that energy was equally partitioned
among collective modes.

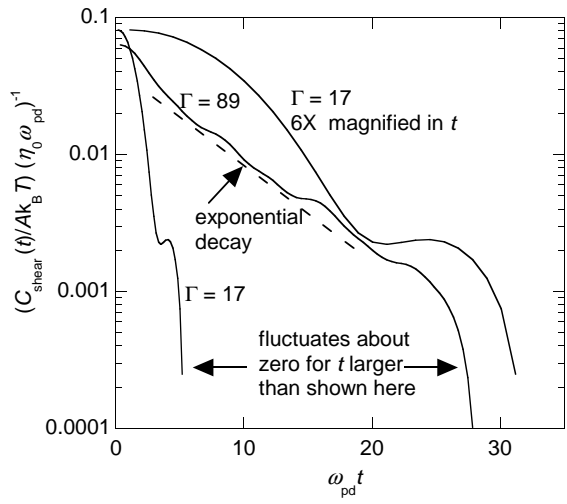
\begin{figure}[p]
\caption{\label{correlation}Shear stress autocorrelation function
$C_{shear}(t)$, computed using Eq.~(\ref{stresscorr}). It is
significant that this function decays exponentially, or even
faster, so that integrating it over time $t$, Eq.~(\ref{Kubo}),
yields a meaningful value of $\eta$.}
\end{figure}

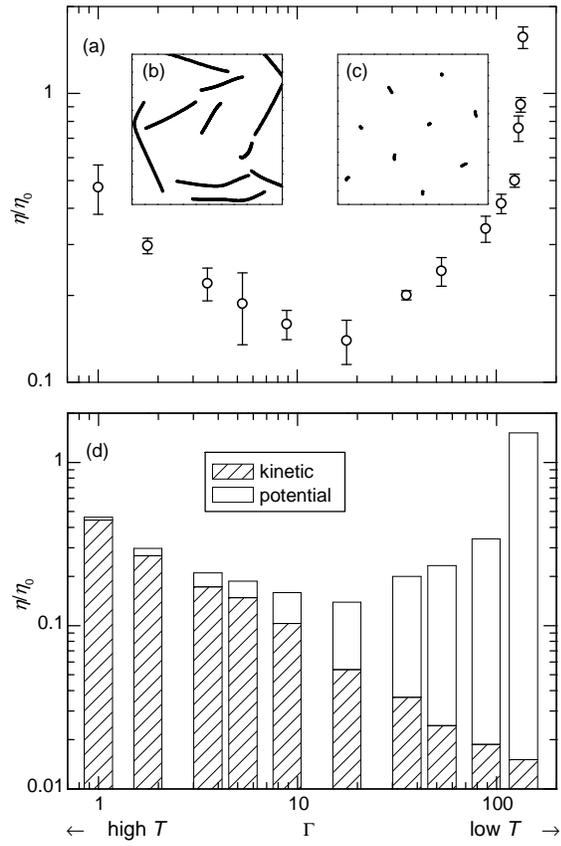
\begin{figure}[p]
\caption{\label{viscosity}(a)~Simulation results show that the 2D
shear viscosity $\eta$ varies with temperature $\Gamma^{-1}$, and
it has a minimum at $\Gamma=17$. Data are shown for $\kappa=0.56$.
Error bars represent the uncertainty. (b)~Trajectories of
particles during a time interval of 1.0 $\omega_{pd}^{-1}$ for
$\Gamma=1$ indicate a state of complete disorder. (c)~Trajectories
of particles during a time interval of 1.0 $\omega_{pd}^{-1}$ for
$\Gamma=124$ indicate a highly ordered structure. (d)~The shear
viscosity $\eta$ is primarily the sum of two contributions
kinetic and potential.}
\end{figure}

Our results were prepared in four steps. First, an initial
configuration of random particle positions and velocities was
chosen. After $T$ reached the desired level and equilibrium was
attained, we began recording data for a duration of
$4\times10^{4}\omega_{pd}^{-1}$. Second, the SACF was calculated
using Eqs.~(\ref{stress}) and (\ref{stresscorr}) for the entire
duration. To verify the validity of this result, we tested ten
ensemble averages, each for a time series of a different duration,
and we found that results for the SACF were independent of the
duration, for the duration we recorded. Third, we integrated the
SACF over $t$, yielding a value for the shear viscosity $\eta$. To
verify that $\eta$ does not depend on $N$, we calculated $\eta$
using $N=4096$ for $\Gamma=17$ and $124$, and we found that $\eta$
was the same, within error bars, as for $N=1024$. Fourth, we
averaged the results for 3 to 6 different initial configurations,
yielding our chief results the SACF, as shown in
Fig.~\ref{correlation}, and $\eta$, as shown in
Fig.~\ref{viscosity}(a).

The SACF decays rapidly with time for $\Gamma<124$. This decay is
almost exponential with $t$ for large $\Gamma$, i.e., $\ln
C_{shear}(t)\propto -t$. It decays even faster, $\ln
C_{shear}(t)\propto -t^{2}$, for small $\Gamma$. These results are
shown for $\Gamma=17$ and 89 in Fig.~\ref{correlation}. Our
results are contrary to some previous results~\cite{Ernst70,
Morriss85} for 2D systems, where a $C_{shear}(t)\propto t^{-1}$
dependence was found in the tail of the stress autocorrelation
function. We will discuss this at the end of this letter.

We find that the shear viscosity calculated using the Green-Kubo
relation is finite in 2D liquids with a Yukawa potential. This is
true for a wide range of $\Gamma$, except possibly near the
freezing region. This is indicated by the exponential (or faster
than exponential) decay of the SACF with time in
Fig.~\ref{correlation}, so that the time integral of the SACF
converges when the Green-Kubo relation is used to calculate
$\eta$. However, when the system is near freezing, i.e.,
$\Gamma>124$, the decay is slower than exponential; thus, our
result for $\eta$ in this regime is less reliable.

Our chief result is the variation of $\eta$ with $\Gamma$,
Fig.~\ref{viscosity}(a). At high temperature, $\eta$ decreases
with $\Gamma$. In this regime, the system behaves as a kinetic
gas, corresponding to a disordered state, as seen from the orbits
in Fig.~\ref{viscosity}(b). When $\Gamma$ is larger than 17, on
the other hand, $\eta$ increases with $\Gamma$; it exhibits an
exponential dependence on $\Gamma$ for $\Gamma<124$ and a much
steeper increase for $\Gamma>124$. Near freezing, $\Gamma>124$,
the system has a highly ordered structure, as seen from the orbits
in Fig.~\ref{viscosity}(c). The minimum viscosity, which is at
$\Gamma=17$, is $0.14\eta_{0}$. Using experimental
parameters~\cite{Vova04} $a=0.6$ mm and $\omega_{pd}=40$
s$^{-1}$, our minimum corresponds to a kinematic viscosity $\nu=2$
mm$^{2}$s$^{-1}$ in physical units. This is about two times larger
than the kinematic viscosity of liquid water at STP conditions.
This result, as noted previously~\cite{Morfill04}, is true even
though $\eta$ is itself an extremely small number for a dusty
plasma. The reason is that the mass density is also very small, so
that the kinematic viscosity, which is the ratio of these two
small quantities, happens to be comparable to that of a denser
substance like water.

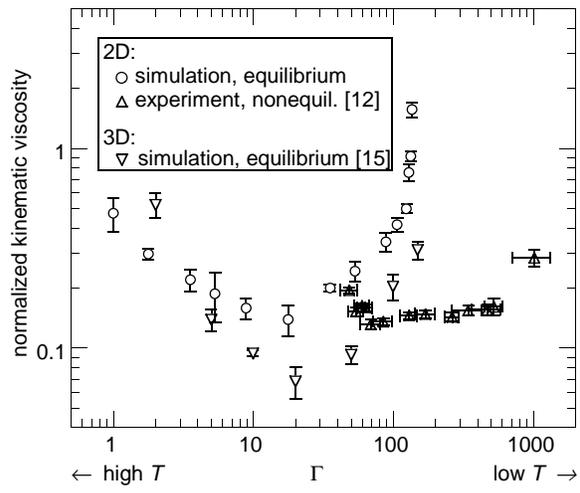
\begin{figure}[p]
\caption{\label{comparison}~Comparison of our 2D simulation with a
2D experiment and a 3D simulation. Nosenko and Goree's
experiment~\cite{Vova04} used a 2D dusty plasma at $\kappa=0.53$
with an externally-applied velocity shear. Saigo and Hamaguchi's
simulation~\cite{Saigo02} used a 3D liquid with a Yukawa
potential at $\kappa=0.50$, in the absence of externally-applied
shear. In all three cases, $\eta$ has a minimum.}
\end{figure}

We compare our results with the experiment of Ref.~\cite{Vova04}
in Fig.~\ref{comparison}. As in our simulation, the experimental
$\eta$ varies with $\Gamma$, and it has a minimum. The minimum
value $\eta=0.13\eta_{0}$ in the experiment matches our result of
$0.14\eta_{0}$.

Aside from this agreement in the magnitude of $\eta$, however,
there is a difference in the value of $\Gamma$ where the minimum
of $\eta$ occurs. The experimental result exhibits a much broader
minimum, and the minimum occurs at a much higher $\Gamma$, as seen
in Fig.~\ref{comparison}. We suggest two reasons for these
differences. Both of these reasons arise from inhomogeneity and
anisotropy in the experiment that are lacking in the simulation.
First, the experiment was nonequilibrium, with an applied shear
that had a specific scale length and that was in a specific
direction. In contrast, the simulation was in equilibrium, with
the shear corresponding to thermal motions that had a wide range
of length scales including very short length scales, and the
direction of the shear fluctuated isotropically. Second, the
experiment had a nonuniform temperature; therefore it had
nonuniform values of $\Gamma$ and $\eta$ whereas the simulation
was uniform. The values reported for $\eta$ and $\Gamma$ in
experiment~\cite{Vova04} were computed as spatial averages over a
region that had a nonuniform temperature; this probably had its
most significant effect on the value of $\Gamma$. Thus, it is not
surprising that the $\Gamma$ for the minimum in the experiment
does not match that of the simulation.

To discover the effect of the dimensionality of a system, we
compare the kinematic viscosity of 2D and 3D liquids. Saigo and
Hamaguchi~\cite{Saigo02} performed an equilibrium simulation
similar to ours, with a Yukawa potential, and their results for
$\kappa=0.5$ are plotted in Fig.~\ref{comparison}. In both cases
the viscosity has a minimum at $\Gamma\approx20$, but the
magnitudes are not the same. In 2D, the kinematic viscosity is
mostly larger for the same value of $\Gamma$.

The minimum of $\eta$ with temperature is a distinctive feature
not found in most simple liquids. In water, for example, viscosity
decreases monotonically with temperature. Systems such as
strongly-coupled plasmas with a long-range repulsive potential,
however, tend to have a minimum. This minimum has been found in a
2D dusty plasma experiment~\cite{Vova04}, simulations of liquids
with Yukawa potential in 2D (the present work) and
3D~\cite{Sanbonmatsu01,Saigo02, Salin02}, and a simulation of a
one-component plasma (OCP)~\cite{Donko00}.

The minimum arises from the temperature dependence of the kinetic
and potential contributions to momentum transport. This is seen in
Fig.~\ref{viscosity}(d), where the kinetic part of $\eta$
decreases with $\Gamma$, while the potential part increases with
$\Gamma$. (There is also a third contribution, called the cross
term~\cite{Hansen86}, but we found it is insignificant for our
conditions.) Neither simple liquids nor dilute gases have a
minimum because they are dominated by the potential and kinetic
contributions, respectively.

Finally, we discuss a controversy for shear viscosity in 2D
liquids. Previous theoretical and simulation efforts, with a
non-Yukawa potential, yielded conflicting results for the decay of
the SACF. This is important because using Green-Kubo relations to
compute transport coefficients requires a decay rapid enough for
the integral to converge. Previous efforts using hydrodynamic
mode-coupling theory~\cite{Ernst70} and an MD
simulation~\cite{Morriss85} predicted a $t^{-1}$ dependence for
the tail in the SACF; this slow decay could lead to a divergent
result for the viscosity. However, other MD simulations
~\cite{Gravina95} yielded a much more rapid decay, which allows
the Green-Kubo integral to converge. Our result for a Yukawa
potential is consistent with the latter, not the former result.
This is true for $\Gamma<124$; for $\Gamma>124$, our data for the
decay was not conclusive, so that further study with a larger
simulation box and more initial conditions is needed to resolve
this controversy in that range, which is very near freezing.

We thank V. Nosenko and F. Skiff for helpful discussions. This
work was supported by NASA and DOE.

\end{document}